\relax
\documentstyle [art12]{article}
\begin {document}
\bibliographystyle {plain}

\title{\bf Existence of Low-temperature Critical  Regime  in a
One-Dimensional Luttinger Liquid with a Weak Link.}
\author {A. M. Tsvelik}
\maketitle
\begin {verse}
$Department~ of~ Physics,~ University~ of~ Oxford,~ 1~ Keble~ Road,$
\\
$Oxford,~OX1~ 3NP,~ UK$\\
\end{verse}
\begin{abstract}
\par
 The exact solution of the boundary sine-Gordon model is
studied in the
region where the scaling dimension of the boundary field $1/2 < \Delta
< 1$. It is shown that at $1 > \Delta \geq 2/3$
the infrared fixed fixed point belongs to the
universality class  of the underscreened anisotrpic Kondo model. The
boundary contribution to the specific heat in this region scales as
$C \sim T^{2\Delta^{-1} - 2}$ at small temperatures.
\end{abstract}

PACS numbers: 72.10.Fk, 72.15.Qm, 73.20. Dx
\sloppy
\par
 The problem of potential scattering in Luttinger liquids has
attracted a great deal of attention since  Kane and Fisher$^1$
mapped it onto the Schmid model$^2$. The latter model is
described by the following action:
\begin{eqnarray}
S = S_0  +
M \int \mbox{d}\tau\cos\left(\beta\phi(0, \tau)/2\right) \nonumber\\
 S_0 = \int \mbox{d}\tau \left\{\int_0^{L}\mbox{d}x \left[
\frac{1}{2}(\partial_{\tau}\phi)^2 +
\frac{1}{2}(\partial_{x}\phi)^2\right]\right\}
\end{eqnarray}
Here
the parameter  $\beta$ is determined by interactions in the bulk.
It is believed that in the region $\Delta = \beta^2/8\pi < 1$ where
the cosine  term is relevant the model scales to the strong coupling
fixed point described by the effective action
\begin{eqnarray}
S_{\mbox{eff}} = S_0 + \frac{1}{2}T_B\int \mbox{d}\tau \phi(0,\tau)^2
\label{eq:eff}
\end{eqnarray}
where
\[
T_B = M(M/\Lambda)^{\Delta/(1 - \Delta)}
\]
and $\Lambda$ is the ultraviolet cut-off.
The robustness of the
effective action (2) has been proven by the instanton
expansion for $\Delta << 1$ $^{1,3}$, by the transformation to the
equivalent free fermion model ($\Delta = 1/2$)$^{4,5}$ and by the exact
solution ($\Delta
\leq 1/2$) $^{6,7}$. In the entire region described by the effective
action (2) the two-point correlation functions on the boundary decay
at T = 0 as $1/\tau^2$ ($\tau >> T_B^{-1}$) and the boundary
contribution to the specific heat is linear in temperature $^{8,9}$.
However, since  all these results have been derived  only  for $\Delta
\leq 1/2$, there is still room for something new in the region
$1/2 < \Delta < 1$.

 In the present Letter we present arguments
that  in the region $2/3 \leq
\Delta < 1$ the model (1) scales to a different critical point where
the $\phi^2$-term does not appear in the low-energy effective action.
In this region the
boundary contribution to the
specific heat   behaves at low temperatures as
\begin{equation}
C_{boundary} \sim  T^{2/\Delta - 2} \: (\Delta > 2/3), \: \: \sim T\ln
T \:
(\Delta = 2/3)
\end{equation}
This behaviour is compatible with the presence on the boundary
of an irrelevant operator with
 scaling dimension $1/\Delta > 1$. The latter is the scaling dimension
of the dual exponent $\cos\left[2\pi\theta(0)/\beta\right]$
( $\theta$ is the field dual to $\phi$).
This  dual exponent is generated by the
instanton expansion of the $\cos(\beta\phi/2)$-potential
and is always present in the low-energy effective
action. However, for $\Delta < 2/3$ its contribution is less important
than that of the
$\phi^2$-term.  On the contrary, in the region $2/3 \leq \Delta < 1$
the instanton processes  completely screen the parabolic
potential. Meanwhile, the operator
$\cos(\beta\phi(0)/2)$ remains  relevant at high energies throughout
the entire region $0 < \Delta < 1$. We suggest that
the low energy fixed point in the  region  $2/3 \leq \Delta < 1$
is described by the effective action
\begin{eqnarray}
S_{\mbox{eff}} = S_0 + \tilde
T_B\int \mbox{d}\tau \cos\left(2\pi\theta(0,\tau)/\beta\right)
\end{eqnarray}
where $\tilde T_B$ is an energy scale related to $T_B$.

 Despite the fact that the boundary sine-Gordon (BSG) model (1)
has been solved exactly in
the sense that exact S-matrices have been found$^6$, complexity of
the solution in the area $\Delta > 1/2$ has prevented it from beeing
studied. At  $\Delta = 1/\nu , \: \nu = 2, 3, ...$ the
thermodynamic Bethe ansatz (TBA) equations were obtained by Fendley {\it et
al.}$^7$:
\begin{eqnarray}
\epsilon_n(v) = \eta\delta_{n,1}\exp(- \pi v/2) \nonumber\\
+ s*\ln\left[1 +
e^{\epsilon_{n - 1}(v)}\right]\left[1 +
e^{\epsilon_{n + 1}(v)}\right] + \delta_{n, \nu - 2}s*\ln\left[1 +
e^{\epsilon_{\nu}(v)}\right]\: (n = 1, ... \nu - 1)\nonumber\\
\epsilon_{\nu}(v) = s*\ln\left[1 +
e^{\epsilon_{\nu - 2}(v)}\right]\label{eps}\\
F_{imp} = - T\int_{-\infty}^{\infty}s[v + \frac{2}{\pi}\ln(T_B/T)]\ln\left[1 +
e^{\epsilon_{\nu - 1}(v)}\right]\label{free}\\
\frac{1}{L}F_{bulk} = - T\int_{-\infty}^{\infty}s[v +
\frac{2}{\pi}\ln(\Lambda/T)]\ln\left[1 +
e^{\epsilon_{1}(v)}\right]
\end{eqnarray}
with $\eta = + 1$. Here
\[
s*f(v) = \int_{-\infty}^{\infty}du\frac{f(u)}{4\cosh[\pi(v - u)/2]}
\]

Eqs.(\ref{eps}) are very similar to  the equations for the
conventional
sine-Gordon model. In the latter case there is a  duality:
 TBA equations are invariant under the transformation
\begin{equation}
\Delta \rightarrow 1 - \Delta = 1/\nu
\end{equation}
except for the free term in the first equation (\ref{eps}), which  changes its
sign: $\eta = - 1$. Therefore  we suggest that TBA equations for
BSG problem at $\Delta = 1 - 1/\nu$ are given by Eqs.(\ref{eps},
\ref{free}) with $\eta = - 1$. We have two additional  arguments
in favour of this proposal. The first one is that the suggested
transformation works for the anisotropic spin-1/2 Heisenberg chain
with open boundary conditions$^8$. It is widely believed that the
latter model adequately describes the strong coupling point of BSG
model. Indeed, at strong coupling the scattering potential becomes
infinite which effectively breaks the chain. The equivalency between
the Luttinger liquid of
spinless fermions and the spin-1/2 Heisenberg chain is established by
the Jordan-Wigner transformation. The second argument is that
at $\Delta = 3/4$
one can derive TBA equations for BSG model
using its  equivalency with the 4-channel anisotropic Kondo model in the
Toulouse
limit$^9$. We shall present the latter argument in detail.

 The Bethe ansatz equations for an
anisotropic k-channel Kondo model are given by
\begin{eqnarray}
[e_k(\eta; u_a)]^Ne_{2S}(\eta; u_a - 1/g) = \prod_{b = 1}^Me_2(\eta; u_a - u_b)
\label{eq}\\
 E = \sum_{a = 1}^M\frac{1}{2\mbox{i}}\ln e_k(\eta; u_a)\\
e_n(\eta; u) = \frac{\sinh[\eta(u - \mbox{i}n)]}{\sinh[\eta(u + \mbox{i}n)]}
\end{eqnarray}
where $S$ is the impurity spin, $g$ is the Kondo coupling constant,
$\eta$ is the anisotropy, N is the
length of the system and $M = kN/2 + S - S^z$ is the number of up
spins. The universal relationship beetween the
quantities $g$ and $\eta$ and the parameters of the Hamiltonian exists
only in the limit of weak anisotropy $\eta << 1$.
The difficulty determining the Toulouse limit is resolved if we  suggest that
this limit
corresponds to the maximal value of $\eta$ at which  the IR fixed point of the
Kondo model still belongs to the same universality class as at $\eta
\rightarrow 0$. The periodicity of the trigonometric factors in
Eqs.(\ref{eq}) suggests that the Toulouse limit is realized at $\eta =
\pi/2(k + 1/\nu)$ ($\nu \rightarrow \infty$). This limit has been
considered in Refs. 10, 11  and the detailed derivation  of TBA
equations is given in Ref. 12. It was shown that TBA equations in the
Toulouse limit have the following form:
\begin{eqnarray}
\epsilon_n = s*\ln\left(1 + \mbox{e}^{\epsilon_{n - 1}}\right)\left(1 +
\mbox{e}^{\epsilon_{n + 1}}\right) +
\delta_{n,k - 2}s*\ln
\left(1 + \mbox{e}^{\epsilon_{k}}\right), \: n = 1, ... k - 1\\
\epsilon_{k} = s*\ln\left(1 + \mbox{e}^{\epsilon_{k - 2}}\right)
- 2\exp( - \pi v/2)\\
F_{bulk} = - NT^2\int_{-\infty}^{\infty} \mbox{d}v \mbox{e}^{ - \pi
v/2}\ln\left(1 +
\mbox{e}^{\epsilon_{k}}\right)\\
F_{boundary} = - T\int_{-\infty}^{\infty} \mbox{d}v s(v + \frac{2}{\pi}\ln
T_B/T)\ln\left(1 +
\mbox{e}^{\epsilon_{2S}}\right)
\end{eqnarray}
where
the temperature $T_B$ is defined as
\[
T_B = \lim_{\nu, \Lambda \rightarrow \infty}\frac{\pi\Lambda}{2\nu}\exp(-
\pi/2g)
\]
Executing this limit one has to be careful to keep the energy scale
$T_B$ finite. Substituting into  these equations  $S = 1/2$ and $k =
4$ we reproduce
the suggested equations for the BSG model with  $\Delta = 3/4$ (i.e. $\nu = 4,
\:
\eta = - 1$). The fact that
this equivalency holds only for $k = 4$ is not suprising. The
conformal charge of the bulk degrees of freedom in the
$k$-channel Kondo model is equal to$^{13,14}$
\begin{equation}
C = \frac{3k}{k + 2}
\end{equation}
At $k = 4$ it is equal to 2 which corresponds to two bosonic modes
interacting with the impurity. In
the Toulouse limit one of these modes decouples from the impurity and
the effective conformal charge becomes 1.

 Now we shall calculate the IR and UV
asymptotics of the boundary free energy for the BSG model with $\Delta  =
1 - 1/\nu$. Analytical solutions are available
for asymptotics of $\epsilon_n(v)$
at $v \rightarrow \pm \infty$ (see, for example, Ref.(15)).
At large temperatures $T >> T_B$ the free energy is determined by the
asymptotics at $v \rightarrow + \infty$ where
\begin{eqnarray}
\left(1 +
\mbox{e}^{\epsilon_{\nu - 1}}\right) = \nu
\end{eqnarray}
so that we have
\begin{equation}
F_{boundary} \rightarrow - \frac{T}{2}\ln\nu
\end{equation}
At small temperatures the leading contribution comes from the region $v
\rightarrow  0$ where $\epsilon_n \: (\nu \neq 1)$ are again almost
constant and $\exp(\epsilon_1)$ is small. Then the
corrections can be determined from the expansion in  $\exp(\epsilon_1)$:
\begin{eqnarray}
g_n(v) \equiv \ln\left(1 + \mbox{e}^{\epsilon_{n}(v)}\right) =
g_n^{(0)} +
g^{(1)}_n(v) + ... \nonumber\\
g_n^{(0)} = 2\ln n \: (n = 2,3, ... \nu - 2), \:
g_{\nu - 1}^{(0)} =  g_{\nu}^{(0)}
\ln(\nu - 1)\nonumber\\
g^{(1)}_n(v) = \frac{1}{2n}[(n + 1) a_n*g_1(v) - (n -
1)a_{n + 2}*g_1(v)]\\
a_n(\omega) = \frac{\sinh(\nu - n)\omega}{\sinh(\nu - 1)\omega}
\end{eqnarray}
Using these expressions we get the following expansion for the free energy:
\begin{eqnarray}
F_{boundary} \rightarrow - \frac{T}{2}\ln(\nu - 1) -
T(\nu - 1)\int_{0}^{\infty} \mbox{d}v
f(v)\ln\left(1 + \mbox{e}^{\epsilon_1(v)}\right) \label{bou}
\end{eqnarray}
where
\[
f(\omega) = \frac{\tanh\omega}{\sinh(\nu - 1)\omega}
\]
The first term in Eq.(\ref{bou}) gives  the finite entropy of the
ground state $S(0) =
- \frac{1}{2}\ln(\Delta^{-1} - 1)$.  A careful analysis shows that this
entropy disappears at $\Delta \leq 1/2$.
The ratio of the partition functions in
the ultraviolet and the infrared limits is
\begin{equation}
Z_{UV}/Z_{IR} = \Delta^{-1/2}
\end{equation}
This result reproduces the expression obtained for $\Delta < 1/2$ in
Ref. 7. The second  term in Eq.(\ref{bou}) gives $\delta F_{imp} \sim
- T^{1 + 2/(\nu - 1)}$ which leads to Eq.(3).

 The author expresses his gratitude to P. Coleman, M. Evans,
E. Fradkin,
L. Ioffe  and P. de Sa for the valuable discussions and the interest to
the work.

\end{document}